\documentclass[a4paper,twoside]{article}

\usepackage{epsfig}
\usepackage{subcaption}
\usepackage{calc}
\usepackage{amssymb}
\usepackage{amstext}
\usepackage{amsmath}
\usepackage{amsthm}
\usepackage{multicol}
\usepackage{pslatex}
\usepackage{apalike}
\usepackage{epstopdf}
\usepackage{SCITEPRESS}     

\def\be{\begin{equation}}
\def\ee{\end{equation}}
\def\miH{\mathbf{H}}
\def\mih{\mathbf{h}}

\def\mix{\mathbf{x}}
\def\miw{\mathbf{w}}
\def\miW{\mathbf{W}}

\def\mi0{\mathbf{0}}

\begin{document}

\title{Towards Terabit LiFi Networking}


\author{\authorname{Ahmad Adnan Qidan, Taisir El-Gorashi1, Jaafar M. H. Elmirghani}
\affiliation{School of Electronic and Electrical Engineering, University of Leeds, LS2 9JT, United Kingdom 
\\Email: \{a.a.qidan, t.e.h.elgorashi, j.m.h.elmirghani\}@leeds.ac.uk}}


\abstract{ Light Fidelity (Li-Fi) is a networked version of optical wireless communication (OWC), which is  a strong candidate to fulfill the unprecedented increase in user-traffic expected in the near future. In OWC, a high number of optical access points (APs) is usually deployed  on the ceiling of an indoor environment to serve multiple users with different demands. Despite the high data rates of OWC networks, due to the use of the optical band for data transmission, they cannot replace current radio frequency (RF) wireless  networks where  OWC has several issues including the small converge area of an optical AP,  the lack of uplink transmission and high blockage probabilities. However, OWC has the potential  to support the requirements in the next generation (6G) of  wireless communications. In this context, heterogeneous optical/RF networks can be considered to overcome the limitations of OWC and RF systems, while providing a high quality of service in terms of achievable data rates and coverage. In this work, infrared lasers,  vertical-cavity surface-emitting(VCSEL) lasers, are used as the key elements of optical APs for serving multiple users. Then,  transmission schemes such  as zero forcing (ZF) and blind interference alignment (BIA) are introduced to manage multi-user interference and maximize the sum rate of users. Moreover, a WiFi system is considered to provide uplink transmission  and serve users that experience a low signal to noise ratio (SNR) from the optical system. To use the resources of the heterogeneous optical/RF network efficiently, we derive  a utility-based objective function that aims to maximize the overall sum rate of the network. This complex problem can be solved using distributed algorithms  to provide sub-optimal solutions with low complexity. The results show that the  sum rate of the proposed hybrid  network is higher than the standalone optical network, using  different transmission schemes.}

\onecolumn \maketitle \normalsize \setcounter{footnote}{0} \vfill

\section{\uppercase{Introduction}}
\label{sec:introduction}

In the last decade, radio frequency (RF) wireless networks have suffered from traffic congestion due to the massive use of the Internet in different fields. Therefore, complementary wireless networks are highly required to offload the traffic of RF wireless networks, while providing high data rates, low latency, low power consumption and  high security, etc. Recently, optical wireless communication (OWC) using license-free optical bandwidth have been investigated to support high communication speeds compared to RF networks. In \cite{Li:14,6736752}, light-emitting diodes (LEDs) are used for illumination and data transmission, achieving high aggregate data rates. Thus,  LED-based OWC systems have low cost infrastructure as sending information occurs while LEDs are already on for providing illumination. In contrast to incandescent bulbs, LEDs are characterized by their small size, long lifetime, high energy efficiency and  low cost. However, LEDs have relatively  low modulation speeds, which may cause limitations in terms of the achievable data rates of OWC systems. In \cite{AA19901111,9566354}, vertical-cavity surface-emitting(VCSEL) lasers are proposed for use as transmitters to serve  users. It is shown that clusters of VCSELs have the potential to provide Tbps data rates due to their high modulation speeds. However, the transmitted power of VCSEL must obey  eye safety regulations.

Interference management is a crucial issue in wireless networks. In particular, an OWC system requires a high number of optical APs to ensure coverage for multiple users distributed in an indoor environment. Therefore, muti-user interference must be addressed to maximize the multiplexing gain of the network. Orthogonal transmission schemes such as  time division multiple access
(TDMA) \cite{8515272}, code division multiple access (CDMA) \cite{8030546} and orthogonal
frequency division multiple access (OFDMA) \cite{8361407} can be implemented for assigning exclusive resources to each user, which results in a low  spectral efficiency. 
 Transmit precoding (TPC) such as minimizing the mean square
error (MMSE) \cite{8119947}, zero-forcing (ZF) \cite{7342273}, or interference alignment (IA) \cite{7893752}, are implemented for optical wireless  networks to serve all users simultaneously. However,  TPC schemes  are derived originally for RF networks, and therefore,  they must be modified  prior  to considering their implementation in  optical wireless networks. In other words, the performance  of TPC schemes is subject to  the unique characteristics of the optical signal. For instance,  the transmitted signal must be strictly non-negative, and therefore, a DC current bias is applied to eliminate the negative values of the transmitted signal, which might cause distortion for useful information. In addition, TPC schemes require channel state information (CSI) at optical transmitters in order to align the interference among users, which is challenging to provide such information in wireless networks \cite{8636954}. 
Recently, blind interference alignment (BIA) has been applied for optical wireless networks to align multi-user interference  without CSI  at transmitters \cite{MMAL17,8636954,9500371}. Particularly, in \cite{MMAL17}, a special optical receiver  called reconfigurable detector is proposed to enable the implementation of BIA in optical wireless networks, where each user must have the ability to switch its channel state following a predefined pattern of BIA. It is shown that BIA overcomes the limitations of the optical signal as its precoding matrix is given by positive values. Besides, BIA avoids the need for  CSI at transmitters, achieving  high user rates compared to other transmission schemes including orthogonal and TPC schemes. However, BIA requires large channel coherence time to guarantee that the  optical channel remains static during the transmission of information to multiple users \cite{5734870}, and therefore, each user can decode its information with minimum error maximizing the DoF of a network. 

A Heterogeneous optical/RF wireless  network can provide a wide  connectivity area for  Mobil users in an indoor environment. It is worth mentioning that each optical AP illuminates a small and confined area referred to as $ attocell $, and  therefore, users might experience high inter-cell interference that results in low SNR. Additionally, uplink transmissions  in OWC systems are usually implemented  using different optical frequency bands. Therefore, hybrid optical/RF networks can overcome these challenges. At this point, resource management and user association must be taken into account  due to the fact that optical and RF systems  differ in terms of coverage and  available resources. 
In \cite{YDAXH}, load balancing is addressed in a hybrid optical/WiFi network to maximize the overall sum rate of the network. In \cite{9064520} an optimization problem is formulated in a BIA-based hybrid optical and WiFi network to relax the limitations of BIA while providing high data rates. In \cite{8417700}, a load balancing approach is proposed, in which a  WiFi system plays the  role of serving users that experience degraded optical system performance. It is worth mentioning that the problems of resource management and user association are usually defined as Mixed integer nonlinear programming (MINLP) problems that have high complexity. In \cite{9521837} distributed algorithms  via Lagrangian multipliers are proposed for solving these problems while  providing sub-optimal solutions with low complexity, where the main problem is divided into sub-problems that can be solved separately. 

In this work, a laser-based optical wireless system  is considered, coexisting of a WiFi system for serving multiple users. We first define our system model, which is composed of multiple optical APs deployed on the ceiling providing downlink transmission and a WiFi AP that provides uplink transmission and  serves users blocked from receiving a high optical power. Then,  the achievable user rate is derived taking  into consideration the implementation of different transmission schemes, ZF and  BIA. Finally, an optimization problem is formulated to manage the resources of the hybrid network jointly with user association. This problem can be solved directly providing an optimal solution with high computational time. Therefore, a distributed algorithm is  proposed to obtain  a solution considerably close to the optimal one with low complexity. The results show that the sum rate of the hybrid network is higher than the standalone optical system.
\begin{figure}[t]
\begin{center}\hspace*{0cm}
\includegraphics[width=1.2\linewidth]{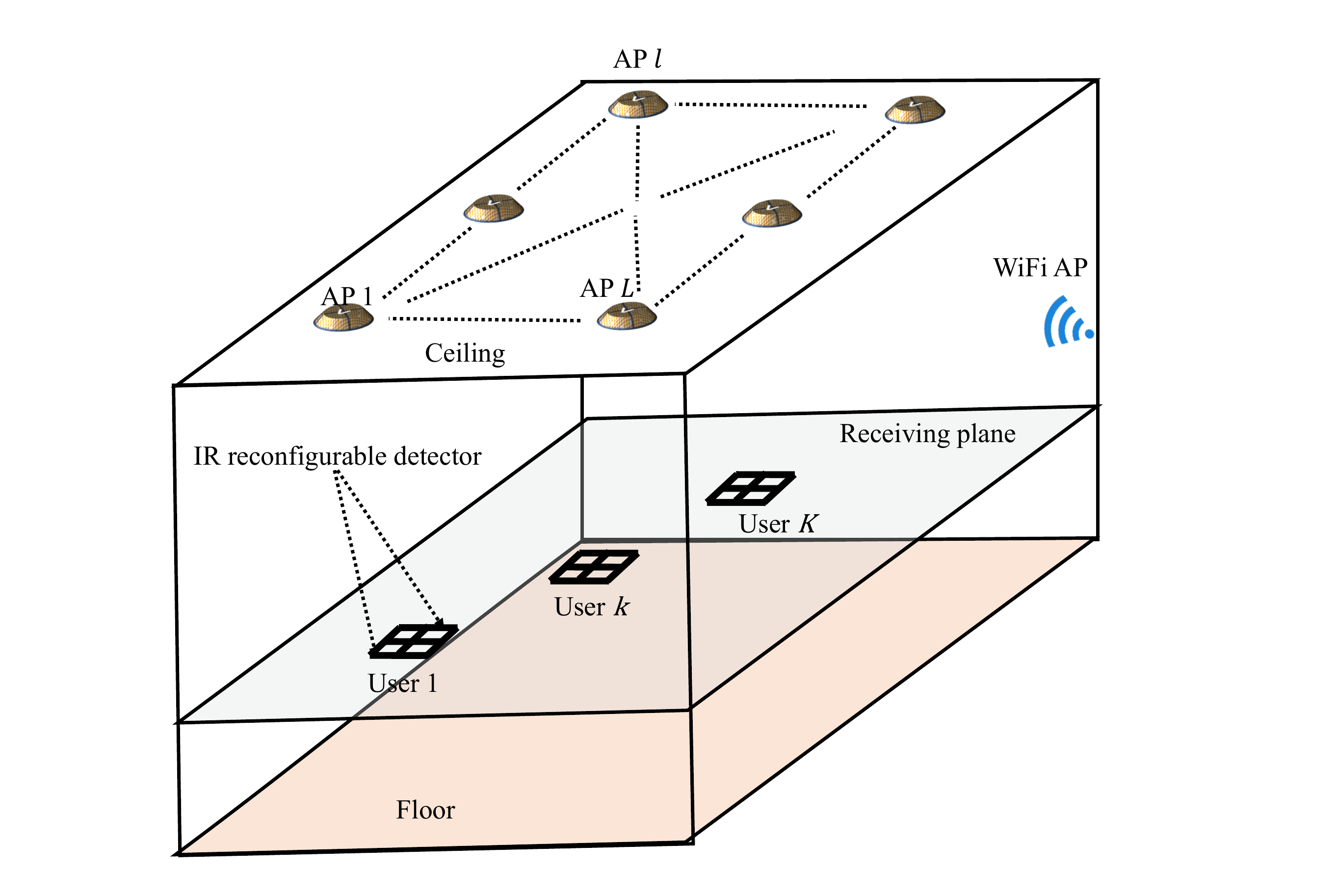}
\end{center}
\vspace{-2mm}
\caption{System model of a hybrid network composed of multiple optical APs and a WiFi system.  }\label{model}
\vspace{-2mm}
\end{figure}

\section{\uppercase{System Model}}

\begin{figure}[t]
\begin{center}\hspace*{0cm}
\includegraphics[width=0.8\linewidth]{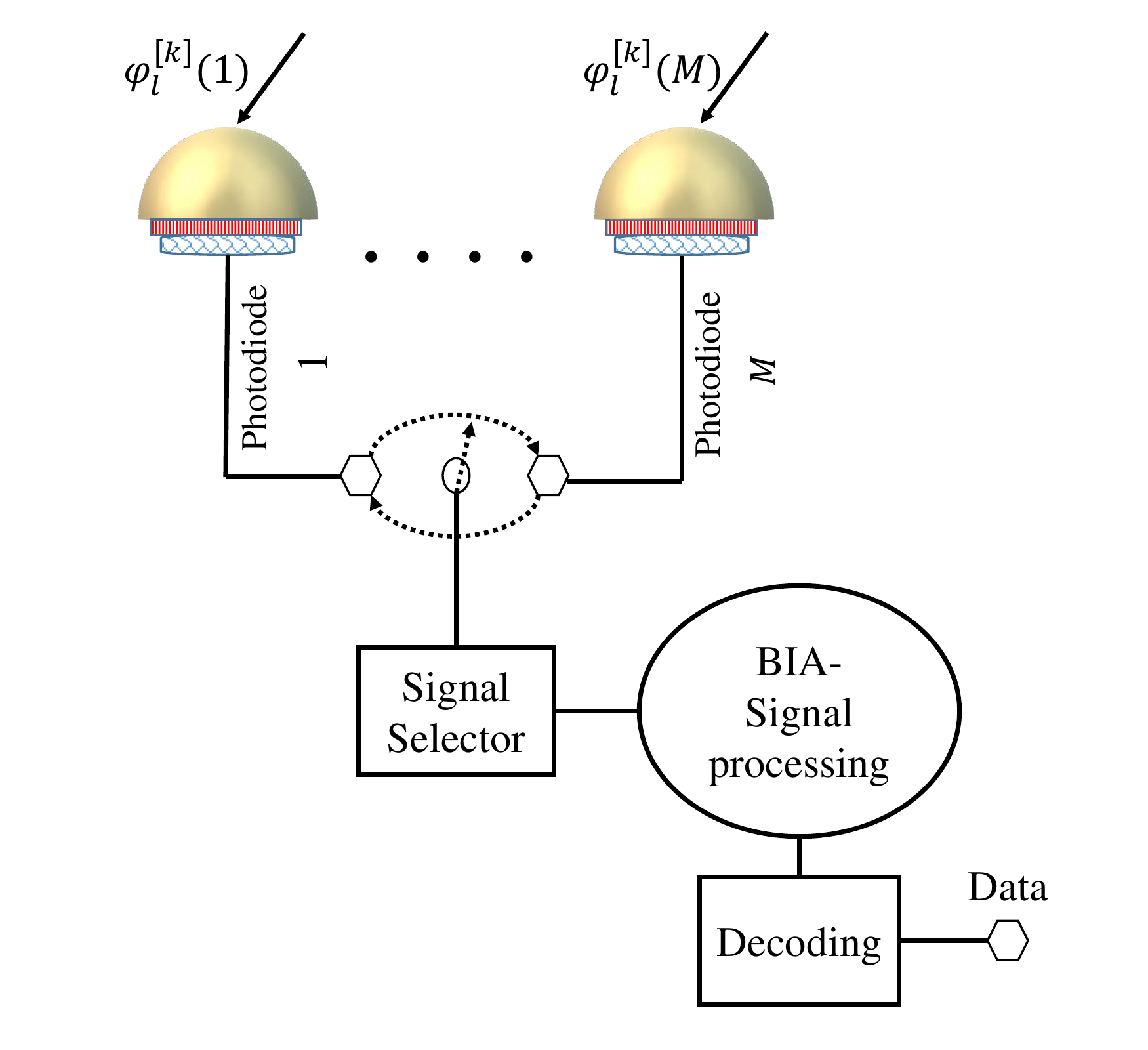}
\end{center}
\vspace{-2mm}
\caption{Reconfigurable optical detector. }\label{dete}
\vspace{-2mm}
\end{figure}

We consider a number  of optical APs deployed on the ceiling given by $ L $, $ l=\{1, \dots, L\} $, and  each optical AP consists  of a certain number of VCSELs, $ L_{v} $,  as shown in Fig. \ref{model}. This optical wireless system serves $ K $, $ k=\{1, \dots, K\} $, users randomly distributed on the receiving plane. To provide a wide field of view (FoV), each user is equipped with a reconfigurable detector composed of $ M $ photodiodes, which point to different directions providing linear independent channel responses, as shown in Fig. \ref{dete}. Focusing on a generic user $ k $, $ k \in K $,  connected to  optical AP $ l $ at time $ n $, the signal received at its photodiode $ m $, $ m \in M $, is

\begin{equation}
y^{[k,l]}[n]=\mih^{[k,l]}(m^{[k,l]}[n])^{T} \mix[n]+  z^{[k,l]}[n],
\end{equation}
where $ \mih^{[k,l]}(m^{[k,l]}[n])^{T} \in \mathbb{R}_+^{L_{v}\times 1} $, $ m^{[k,l]}[n] $  is the mode of photodiode $ m $ at time slot $ n $, $ \mix $ is the transmitted signal and $ z^{[k,l]} $ is real valued additive white Gaussian noise with zero mean and variance given by the sum of  shot noise, thermal noise and the intensity noise of the laser, i.e., 
\begin{equation}
 \sigma^{2}_{z}=\sigma_{\sum}+\mathrm{RIN} \left(\sum^{L}_{l', l'\neq l} ~( \mih^{[k,l']} ~ \delta_{m}~ P_{t,l'} )^2\right)~B, 
\end{equation}
where $\sigma_{\sum} $  is the sum of shot noise and thermal noise, and $ \mathrm{RIN} $ is  the relative intensity noise of the VCSEL transmitter \cite{AA19901111}. Note that, $ \mih^{[k,l']} $ is the interference channel between user $  k$ and  adjacent optical AP $ l' $, $ \delta_{m} $ is the responsivity of photodiode $ m $, $ P_{t,l'} $ is the optical power of AP $ l' $ and $ B $ is the single-sided bandwidth of the system.

In this work, a WiFi AP is considered to provide uplink transmission and serve users that receive low optical power from  the optical APs. In this context, the optical and WiFi systems  are connected to a central unit that controls user association and  the resources of the network such that load balancing between the systems is achieved \cite{9064520}. Moreover, the central unit has information on the distribution of the users, in addition to some information that varies based on the transmission scheme considered for managing the interference among the users in the hybrid network. In particular, orthogonal transmission is considered for the WiFi system, while ZF or BIA  is applied for the optical system.     

 
\subsection{Optical transmitter}

The optical channel  is usually given by Line-of-Sight (LoS) and diffuse components. That is, the  optical channel between  AP $ l $ and user $ k $ at photodiode $ m $ is    
\begin{equation}
h^{[k,l]}(m)=h_{ \mathrm{LoS}}^{[k,l]}(m) + h_{\mathrm{diff}}(f) e^{-j2\pi f \Delta T}, 
\end{equation}
where $h_{ \mathrm{LoS}}^{[k,l]}(m)$ corresponds to the LoS link, $h_{\mathrm{\mathrm{diff}}}$ are non-LoS components and $\Delta T$ is the delay between them. Note that,  a reconfigurable detector is considered in this work, and therefore, each user has  a wide FoV receiver, which results in detecting a large portion of the optical power radiated from LoS links. Given this point, non-LoS components can be  neglected for the sake of simplicity \cite{AA19901111}.

The power distribution  of the  VCSEL-based optical AP follows a Gaussian function of multiple modes. In this context, the transmitted power of the optical AP  is determined by the beam waist $ W_{0} $, the wavelength $ \lambda $ and the distance $ d $ from the ceiling to the receiving plane. Thus, the beam radius of  the laser at distance $  d $ is 
\begin{equation}
W_{d}=W_{0} \left( 1+ \left(\frac{d}{d_{Ra}}\right)^{2}\right)^{1/2},
\end{equation}
where $ d_{Ra} $ is the Rayleigh range given by 
\begin{equation}
d_{Ra}= \frac{\pi W^{2}_{0} n }{ \lambda},
\end{equation}
where $ n=1 $ is the refractive index of air. Moreover, the  intensity spatial distribution  of the laser over the transverse plane at distance $ d $ can be expressed as  

\begin{equation}
I(r,d) = \frac{2 P_{t}}{\pi W^{2}_{d}}~ \mathrm{exp}\left(-\frac{2 r^{2}}{W^{2}_{d}}\right), 
\end{equation}
where $ r $ is the radial distance from the center of the beam spot and the distance $ d $.

Considering that the  detection area of the reconfigurable detector is $ A_{rec} $,  all the photodiodes of the reconfigurable detector have  the same size and detection area. Thus, the detection area of  photodiode $ m $ is given by $ A_m = \frac{A_{rec}}{M} $, $ m \in M $. In this case, the received power  of user $ k $ at  photodiode $ m $ from  optical AP  $ l $ is given by 
\begin{equation}
\begin{split}
&P_{m,l}=
\int_{0}^{A_m /2 \pi} I(r,d) 2\pi r dr \\ 
&= P_{t,l}\left[1- \mathrm{exp}\left(- 2 \left(\frac{ A_{m}}{2 \pi W_{d}}\right)^{2}\right)\right],
\end{split}
\end{equation}
\subsection{Reconfigurable detector}
The receiver considered in this work is composed of $ M $ photodiodes that provide channel responses linearly independent following an angle diversity arrangement \cite{MMAL17}. As shown in Fig. \ref{dete}, all the  photodiodes are connected to a single signal processing chain through a selector, minimizing the power consumption of the receiver. Each photodiode $m$ has a unique orientation
given by its polar and azimuth angles, which are  denoted as $ \theta^{[k,m]}$ and $\alpha^{[k,m]}$, respectively. Thus, the orientation vector of photodiode  $m$  of user $k$ is  given by
 \begin{equation}
\begin{split}
\hat{\mathbf{n}}^{[k,m]}=&
\left[ \sin\left(\theta^{[k,m]} \right)\cos\left(\alpha^{[k,m]}\right), \,\, \right . \\
& \phantom{\{} \left . \sin\left(\theta^{[k,m]} \right)\sin\left(\alpha^{[k,m]}\right), \,\, \cos\left(\theta^{[k,m]} \right) \right],
\end{split}
\end{equation}

Moreover,  the irradiance and incidence angles are determined by
\begin{equation}
\phi_{l}^{[k]}=\arccos\left(\frac{\hat{\mathbf{n}}_{l} \cdot \mathbf{v}_{l}^{[k]}}{\Vert \hat{\mathbf{n}}_{l} \Vert \Vert \mathbf{v}_{l}^{[k]} \Vert} \right)
\end{equation}
\begin{equation}
\varphi_{l}^{[k]}(m)=\arccos\left(\frac{\hat{\mathbf{n}}^{[k,m]} \cdot \mathbf{v}_{l}^{[k]}}{\Vert \hat{\mathbf{n}}^{[k,m]} \Vert \Vert \mathbf{v}_{l}^{[k]} \Vert}\right)
\end{equation}
respectively, where $\hat{\mathbf{n}}_{l}$ is the normal orientation vector of  optical AP $l$ and $\mathbf{v}_{l}^{[k]}$ is the vector from  optical AP $l$ to  user $k$. Note that, each optical AP points to  the floor, and hence, the normal orientation vector is $\hat{\mathbf{n}} = [0, 0, -1]$. It is worth mentioning that the photodiodes arrangement of a reconfigurable detector can follow 
different geometrical patterns including  pyramidal, hemispherical or random receiving orientation angle (ROA) distributions such that providing distinct and linearly independent channel responses is guaranteed.

At this point, the channel matrix of user $ k $ connected for example to $ L $ optical APs is given by
\begin{equation}
\begin{split}
&\mathbf{H}^{[k]} =\\
&
\begin{bmatrix}
h_{1}^{[k]}(1)^{T} & h_{2}^{[k]}(1)^{T} & \ldots & h_{L}^{[k]}(1)^{T} \\
h_{1}^{[k]}(2)^{T} & h_{2}^{[k]}(2)^{T} & \ldots & h_{L}^{[k]}(2)^{T} \\
\vdots &   &   & \vdots \\ \\
h_{1}^{[k]}(M)^{T} & h_{2}^{[k]}(M)^{T} & \ldots & h_{L}^{[k]}(M)^{T} \\
\end{bmatrix},
\end{split}
\end{equation}
where $  h_{l}^{[k]}(m)$  is the channel response between optical AP $l$ and user $ k $ at 
the photodiode $m$ of the reconfigurable detector. Note that, the reconfigurable detector  must consist of at least $ M=L $ photodiodes in order to ensure that $\mathbf{H}^{[k]}$ is a full-rank matrix. 
\subsection{Uplink transmission}
A WiFi system is considered in the hybrid network to provide uplink transmission. In this context, each user has an RF interface to communicate with the WiFi system, and the power allocated to that RF interface is denoted by $ P_{\mathrm{up}} $. The WiFi system devotes $ e_{\mathrm{up}} $ resources  for the uplink  transmission, allocating $ e^{[k]}_{\mathrm{up}}= e_{\mathrm{up}}/K $ resources for each user. Note that, the power consumption at the RF interface of each user must be taken into account due to battery limitations. Hence, denoting $ \rho $  as power amplifier efficiency,  the power consumption equals to $ P_{\mathrm{up}} /\rho $. It is worth pointing out that ensuring power consumption within the limit requires the definition of three constraints for the maximum power allocated to the RF interface of the user,  the battery power constraint of the user and the minimum rate of the uplink transmission.

 
\section{\uppercase{Multiple access schemes}}
In general,  optical wireless systems are composed of a high number of optical transmitters that serve multiple users. In such high density wireless networks,  multi-user interference is a crucial issue that must be managed efficiently to achieve a high spectral efficiency. In this context, low complexity orthogonal transmission schemes  such as TDMA \cite{8515272} and  OFDMA \cite{8361407} are implemented for this purpose where exclusive time or frequency slots are allocated to each user. In addition, WDMA can be implemented for avoiding multi-user interference by allocating a certain wavelength to each user  while using optical filters on the user side to eliminate the other wavelengths. Despite the low complexity of the orthogonal transmission schemes, the utilization of the network resources might be minimized. In the following, advanced transmission schemes are introduced to serve multiple users simultaneously. Note that, the equations below are derived for a simple full connectivity optical wireless system where $ L $ APs serve $ K $ users. However, they can be simply extended for our system model in Section 2. 
       
\subsubsection{Precoding transmission schemes}
TPC schemes, such as ZF in  \cite{7342273}, are proposed for interference management in optical wireless networks assuming  prefect CSI at transmitters as well as  cooperation among optical APs. 

Let us consider  linear transmit precoding to manage muti-user interference, and then, derive the achievable user rate. Focussing on user $ k $, its  precoding vector is  denoted as $\miw^{[k]}  \in \mathbb{R}_+^{L\times 1}  $, and its received signal is given by

\begin{equation}
\label{zf}
y^{[k]}=\mih^{[k]} \miw^{[k]} s^{[k]}+ \mih^{[k]} \sum^{K}_{k', k'\neq k} \miw^{[k']} s^{[k']}+  z^{[k]},
\end{equation}
where the first term is the useful information intended to user $ k $ and the second term is the interference  received due to transmission to the other users, $ k'\neq k $. Moreover,  $ s^{[k]} $ is the symbol transmitted to user $ k $. Note that, On–off keying (OOK)  can be used in optical wireless communication for the sake of avoiding complexity. The interference term in equation \eqref{zf}  can be  canceled using the precoding of the ZF scheme.  Furthermore, the channel matrix for the scenario considered in this section, $ K $ users connected to $ L $ AP, is given by 

\begin{equation}
\miH=\left[ \mih^{[1]}, \dots, \mih^{[k]}, \dots, \mih^{[K]}\right]^{T}, 
\end{equation} 
where $ \miH  \in \mathbb{R}_+^{K\times L}  $, and  the precoding matrix is given by 
\begin{equation}
 \miW=\left[ \miw^{[1]}, \dots, \miw^{[k]}, \dots, \miw^{[K]}\right],
\end{equation}
where $\miW \in \mathbb{R}_+^{L\times K} $. Thus, $ \miH \miW= \mathrm{diag} (\sqrt{g_{k}})$, where $ g_{k} $ is the channel gain of user $ k $ after the ZF precoding. In \cite{7342273}, the lower band  user capacity is derived taking into consideration the implementation of the ZF scheme. It is expressed as 

\begin{equation}
C_{\mathrm{zf}}^{[k]}\geq \frac{1}{2} \log \left( 1+ \frac{ 2 \mid \mih^{[k]} \miw^{[k]} \mid^{2} }{\pi e \left(\sum^{K}_{k', k'\neq k}   \frac{1}{3}\mid \mih^{[k]} \miw^{[k']} \mid^{2} + \sigma^{2}_{z}\right)}\right).
\end{equation} 
Therefore, the achievable user rate of ZF, $ r_{\mathrm{zf}}^{[k]}$,  can be written as follows  
\begin{equation}
\label{zfop}
r_{\mathrm{zf}}^{[k]}= \frac{1}{2} \log \left( 1+ \frac{ 2 \mid \mih^{[k]} \miw^{[k]} \mid^{2} }{\pi e \left(\sum^{K}_{k', k'\neq k}   \frac{1}{3}\mid \mih^{[k]} \miw^{[k']} \mid^{2} + \sigma^{2}_{z}\right)}\right),
\end{equation}
and the sum rate is given by $ R_{\mathrm{zf}}= \sum^{K}_{k=1} r^{[k]} $. It is worth mentioning that the pseudo-inverse is
$ \miH^{\dagger}= \miH^{H} (\miH \miH^{H})^{-1} $, which  obeys
the ZF criterion to  obtain interference-free signals at users.

In addition to  the  requirement of TPC schemes in terms of  the need for CSI at transmitters, 
the  use of TPC schemes in optical wireless networks is  subject to limitations due to the optical signal characteristics such as the non-negativity of the transmitted signal and the high correlation among the channel responses of users \cite{MMAL17,8636954,9500371}. It is shown that these schemes achieve low data rates  compared to  their performance in RF networks. Therefore, the following points must be taken into consideration prior to considering TPC schemes for interference management.

\begin{itemize}
    \item  The need for prefect CSI at optical transmitters.
 \item Cooperation among optical APs is required.
 \item The precoding matrix for determining the transmitted signal contains
negative and positive values, and therefore,  a DC bias current must be applied to guarantee the non negativity of the transmitted signal.
 \item The performance of TPC schemes is subject to  correlation among the
channel responses of users.
\end{itemize}

\begin{figure*}[t]
\begin{center}\hspace*{0cm}
\includegraphics[width=0.75\linewidth]{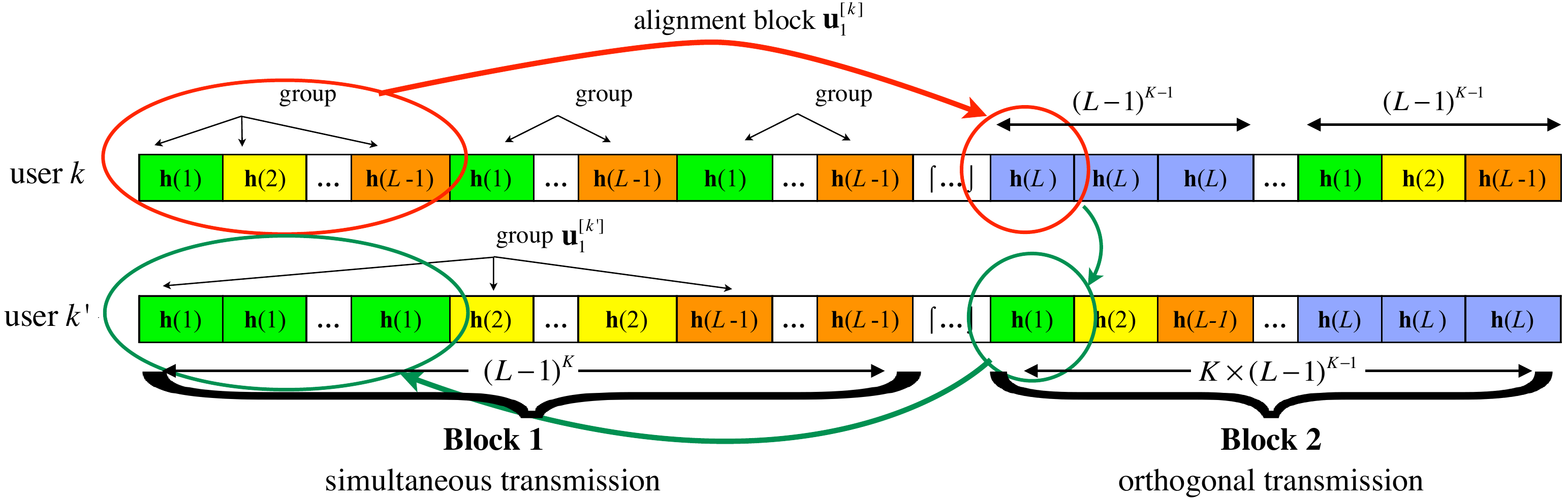}
\end{center}
\vspace{-2mm}
\caption{The transmission block of BIA. Each color represents a preset mode. }\label{BIA}
\vspace{-2mm}
\end{figure*}
\subsubsection{Blind interference alignment}
Following \cite{5734870}, in this section, the methodology of BIA is introduced in detail in a multiple-input multiple-output broadcast channel (MISO BC) scenario where $ L $ APs serve $ K $ users.
 The key idea of  BIA is  aligning  the useful information  intended to each user into a full rank matrix,  while having  the interfering signals contained  into a matrix that has  at least one dimension less. Basically, BIA allocates  a number of alignment blocks determined in accordance with the size of the network, i.e., the number of transmitters and users, to each user simultaneously. During an alignment block of user $k$, $ k \in K $,   the channel state of that user changes among $L$ distinct preset modes, while the channel states of all other users remain constant, as shown in Fig. \ref{BIA}. Note that, the reconfigurable photodetector derived in Section 2 gives the user  the ability to vary its channel state among the symbol extensions, i.e., the time slots, of its alignment block. Denoting a symbol transmitted to user $ k $ as $ \mathbf{u}_\ell^{[k]} $, where $\ell$ th is the index of an alignment block allocated to user $k$, BIA guarantees the linear independence of the symbols $\mathbf{u}_\ell^{[k]}=\{{u}_{\ell,1}^{[k]}, \dots,{u}_{\ell,L}^{[k]}\} $ received  from $ L $ transmitters  over alignment block $\ell$. Additionally, the independence among  multiple alignment blocks allocated to each user must be ensured   during the entire transmission block. To satisfy this condition, the alignment blocks of each user must be transmitted in orthogonal fashion such that any pair of  alignment blocks allocated to the same user do not contain any symbol in common, giving each user the ability to decode the information transmitted  over its alignment blocks.     
 
Once the independence among the desired symbols of  user $k$ is ensured, the interference received due to  transmission to all other users,  for example  $\mathbf{u}_\ell^{[k']}$ transmitted to user $ k' $, $k' \neq k$, can be aligned into a matrix that has  less dimensions than the data streams $\mathbf{u}_\ell^{[k]}$ intended to user $k$ during each of its alignment blocks. Given this point, the interference can be measured and subtracted, and  $L$ DoF carried by $\mathbf{u}_\ell^{[k]} $ can be decoded by user $k$.

To guarantee the methodology mentioned above, the transmission block of BIA is divided into Block 1 and Block 2, comprising  $ (L-1)^{K}+ K (L-1)^K-1 $ time slots in total. As a result, $ (L-1)^{K-1}$ alignment blocks are  allocated to each user. Referring to Fig. \ref{BIA}, the contractions of Block 1 and Block 2, as well as the length of each block, are  described as  follows.  In  Block 1, transmissions to all users occur simultaneously, causing severe  interference among them. On the other hand, each user is served in orthogonal fashion over  Block 2 in order to  give users enough dimensions to measure and cancel the interference received during Block 1, i.e., user $k$ measures the symbols transmitted to all other users $\mathbf{u}_\ell^{[k']} $, $ k' \neq k$, during Block 2, and subtracts it afterwards from the signal received at Block 1. Therefore, the first $(L-1)$ slots of each alignment block  must belong to Block 1 forming a group, while the last  time slot of each alignment block is provided over Block 2. As a result, the length of Block 1 is given by $ (L-1)^{K} $ time slots, while Block 2 comprises $ K (L-1)^{K-1} $ time slots. 

At this point,  the signal received by user $k$ from $ L $ transmitters during an alignment block after interference subtraction can be written as
\begin{equation}
\label{eq:rate_vlc_C}
\tilde{\mathbf{y}}^{[k]} = \mathbf{H}^{[k]}\mathbf{u}_{\ell}^{[k]}      +  \tilde{\mathbf{z}}^{[k]},
\end{equation}
where $\tilde{\mathbf{y}}^{[k]} \in \mathbb{R}^{L \times 1}$ is the signal received during the $L$ time slots of alignment block $ \ell $.  Moreover, in~\eqref{eq:rate_vlc_C}, 
\begin{equation}
\mathbf{H}^{[k]} = \begin{bmatrix} \mathbf{h}^{[k]}(1) & \dots & \mathbf{h}^{[k]}(L)
 \end{bmatrix} \in \mathbb{R}_+^{L\times 1},
\end{equation}
is the channel matrix of user $k$ that contains $L$ linearly independent channel responses, i.e., $\mathbf{H}^{[k]} $ is a full rank matrix, and $\tilde{\mathbf{z}}^{[k,c]}$ is real valued
additive white Gaussian noise with zero mean and variance given by the sum of thermal noise, shot noise and noise that results from interference subtraction, which is defined as a covariance matrix as follows
\begin{equation}
\label{eq:noiseBIA_cell}
\mathbf{R}_{\tilde{z}}
= 
\begin{bmatrix}
K \mathbf{I}_{L-1} & \mathbf{0}_{L-1,1}\\
\mathbf{0}_{1,L-1} & 1
\end{bmatrix}.
\end{equation}
It is worth mentioning that the performance of BIA is limited in a high density network due to  increase in subtraction noise as  the number of users increases. Moreover,  the requirements of large  channel coherence time become more  difficult to guarantee in such high density networks.   
In BIA, the achievable user rate  can be written as 
\begin{equation}
\begin{split}
\label{VLC_r}
r_{\mathrm{bia}}^{[k]} =& e_{\mathrm{bia}}^{[k]}\\
&\times \mathbb{E}\left[\log_2\det\left(\mathbf{I}_L+P_{\mathrm{str}}\mathbf{H}^{[k]}{\mathbf{H}^{[k]}}^{H}{\mathbf{R}_{\mathsf{z_I}}^{[k]}}^{-1} \right)\right],
\end{split}
\end{equation}
where $ e_{\mathrm{bia}}^{[k]}= \frac{1}{L+K -1}  $ is the ratio of the number of alignment blocks allocated to user $k$ to the length of the BIA transmission block, $ P_{\mathrm{str}}$ is the power allocated to the data stream and $ \mathbf{R}_{\mathsf{z_I}}^{[k]} $ is the covariance matrix of noise.



\begin{table}
\centering
\caption{Simulation Parameters}
\begin{tabular}{|l|c|}
\hline
{\bf Optical system parameters}	& {\bf Value} \\ \hline\hline
Bandwidth of VCSEL laser	& 5 GHz \\\hline
Wavelength of VCSEL laser	& 830 nm \\\hline
 VCSEL beam waist	& $ 5-30~ \mu $m \\\hline
Physical area of the photodiode	&15 $\text{mm}^2$ \\\hline
Receiver FOV	& 45 deg \\\hline
Detector responsivity 	& 0.53 A/W \\\hline
Gain of optical filter & 	1.0 \\\hline
Laser noise	& $-155~ dB/H$z  \\\hline\hline
{\bf WiFi parameters}	& {\bf Value} \\ \hline\hline
OFDM subcarrier number                    &   108   \\\hline
Transmitted power for Wi-Fi  AP	& 20 dBm \\\hline
Bandwidth for WiFi  AP	 &  40 MHz \\\hline
Noise power of WiFi	& -63 dBm \\\hline
\end{tabular}
\end{table}

\section{\uppercase{Optimization problem}}
The resources of the optical wireless system must be managed among users taking into consideration the existence  of the complementary WiFi system to maximize the sum rate of the hybrid optical/RF network. Specifically, an optimization problem is formulated in this section with an objective function that jointly finds the optimal user assignment and resource allocation for multiple users. In previous works on heterogeneous networks, SNR maximization is considered for user assignment to avoid complexity where users simply connect to a wireless system that provides high SNR. However, this approach is not efficient when optical and RF systems  work together due to the fact that an optical AP illuminates a small area compared to an RF AP, and therefore, considering SNR maximization might lead to overloading one of these systems \cite{9064520}. Consequently, resource management in such  hybrid optical/RF networks must be based on rate maximization where users connect to the system that has more available resources than the other.  

%
%
Let us define a new notation $ \mathcal{L}, l= \{1, \dots, \mathcal{L} \} $, that contains all the available APs, i.e., optical and WiFi APs, in the indoor environment, as shown in Fig \ref{model}. Moreover, an expression for the achievable user-rate regardless of the optical and WiFi systems can be written as  


\begin{equation}
\label{r_uti}
R^{[k,l]}=e^{[k,l]} r^{[k,l]},
\end{equation}
where {$r^{[k,l]}$ is the  achievable user rate of a generic user $k$  connected to optical or WiFi APs, and $e^{[k,l]}$ denotes the fraction of the resources  allocated to user $ k $ from AP $ l $. Note that, the achievable user rate for the optical system is given by equations \eqref{VLC_r} or \eqref{zfop}  based on the transmission scheme considered, BIA or ZF, respectively. On the other hand, the achievable user rate of the WiFi system can be easily derived considering orthogonal frequency-division multiplexing (OFDM) with multiple subcarriers, which are allocated in orthogonal fashion to the users connected to the WiFi system avoiding multi-user interference as in \cite{9064520}.}

The overall achievable rate of a generic user $k$ is given by
\begin{equation}
\label{r_uti2}
R^{[k]}= \sum\limits_{l \in \mathcal{L}}  x^{[k,l]} e^{[k,l]} r^{[k,l]},
\end{equation}
where  $x^{[k,l]}$ is a  variable that determines user association. In this context,  the variable  $x^{[k,c]}=1$ if user $k$ is connected to AP $l$, otherwise the variable $ x^{[k,c]}=0$.

At this point, our aim is to formulate an optimization problem  that maximizes the sum  rate of all the users by assigning each user to an AP that provides a high user-rate.  In  this context, we define a utility function of the overall user rate in equation \eqref{r_uti2} under several constraints that ensure the maximization of the hybrid network overall sum rate. That is  
%
\begin{align}
\max_{x,e}. & \sum\limits_{k\in  K}   \varphi \left( \sum\limits_{l \in \mathcal{L}} x^{[k,l]} e^{[k,l]} r^{[k,l]} \right)\label{eq:ML} \\
\textrm{s.t.} \quad & \,\sum\limits_{l \in \mathcal{L}}x^{[k,l]} = 1 ~~~~~~\forall k\in K \nonumber\\
\quad & \sum\limits_{k\in K} e^{[k,l]} \leq 1 ~~~~~~\forall l \in \mathcal{L}\nonumber\\
& \,\, 0 \leq e^{[k,l]} \leq 1,\,  x^{[k,l]} \in \left\{0,1\right\},\,\,  \forall l \in \mathcal{L} , \forall k\in  K, \nonumber
\end{align}
where $ \varphi (.)$  is a  strictly concave function that  achieves some levels of fairness among the users of the network if it is considered in its logarithmic form, i.e., $ \varphi (.)= \log (.)$ \cite{9064520,9521837}. 
The first constraint in~\eqref{eq:ML} ensures  that each user connects to only one AP, i.e.,  one of the optical APs or the WiFi system.  The second constraint guarantees that the portion of the transmission resources employed for each optical AP or the WiFi system is less than 1. Moreover, the last constraint considers the feasible region of the optimization variables, where $x^{[k,l]}$ and  $e^{[k,l]}$ are binary and real variables between 0 and 1, respectively.
\begin{figure*}[t]
\begin{center}\hspace*{0cm}
\includegraphics[width=0.65\linewidth]{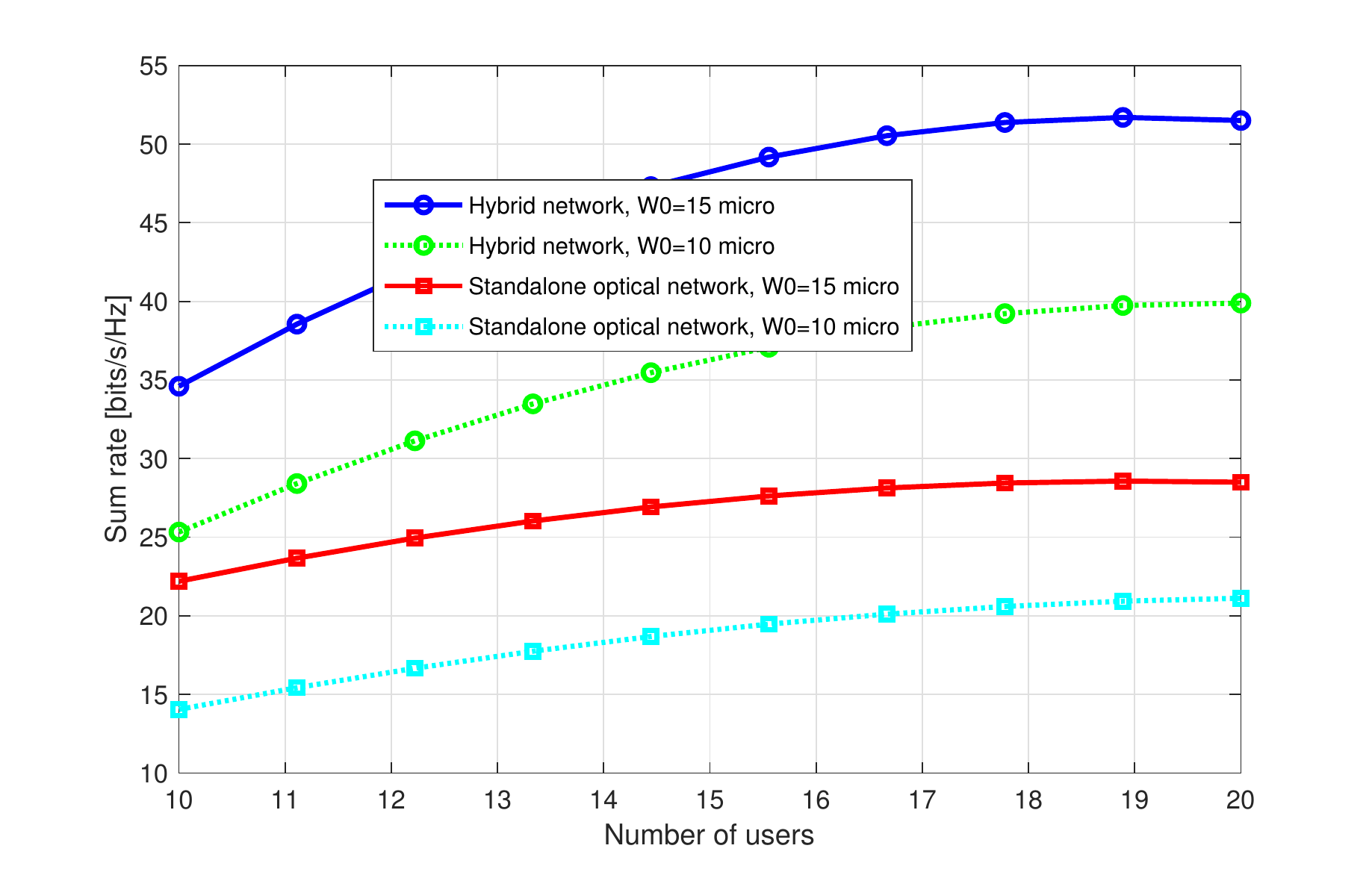}
\end{center}
\vspace{-2mm}
\caption{Sum rates of hybrid optical/RF and standalone optical networks versus the number of users, considering BIA for data transmission.}\label{rate_users}
\vspace{-2mm}
\end{figure*}

\begin{figure*}[t]
\begin{center}\hspace*{0cm}
\includegraphics[width=0.65\linewidth]{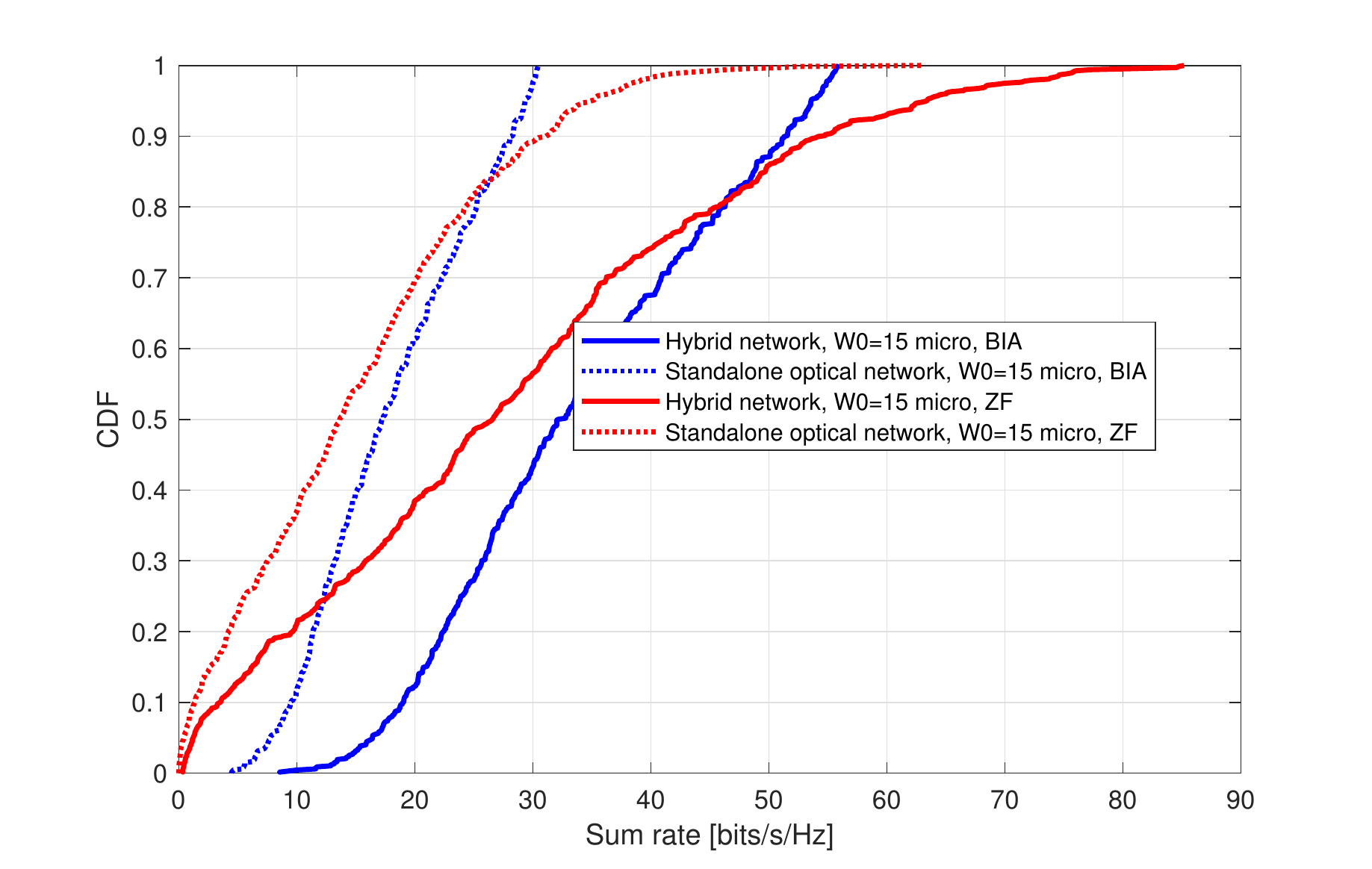}
\end{center}
\vspace{-2mm}
\caption{ CDF of the sum rate for hybrid optical/RF and standalone optical networks.  }\label{cdf}
\vspace{-2mm}
\end{figure*}
The problem in \eqref{eq:ML} is MINLP in which  two variables, $x^{[k,l]}$ and  $e^{[k,l]}$,  are coupled, and therefore, it is not easy to solve. In this sense, a decentralized algorithm via Lagrangian multipliers can be used to provide an effective solution with low computational time. In particular, the Lagrangian function of the main problem in  \eqref{eq:ML} is formulated as 
\begin{equation}
\label{Lag}
\begin{split}
&f(x, k,\mu) =\\ 
&\underbrace{\sum\limits_{k\in k} \sum\limits_{l \in \mathcal{L}} x^{[k,l]}\left(\log (r^{[k,l]})- \mu_{l}\right)}_{s(x , \mu_l)}
 + \underbrace{\sum\limits_{l \in \mathcal{L}} K_{l} \left( \mu_{l}-\log(K_{l})\right)}_{g(k, \mu_l)},
\end{split} 
\end{equation}
considering $ \sum_{k \in K}x^{[k,l]}=K_{l}$ is  the load on AP $ l $ and uniform resource allocation  among $ K_{l} $ users, i.e., $ x^{[k,l]}= \frac{1}{K_{l}}  $. Note that, $\mu_{l}$ is a   Lagrangian multiplier that works as a message between  AP $l $ and users sending requests to connect to that AP. The  problem in \eqref{Lag}  can be divided into sub-problems, $ s(x , \mu_l) $ and $ g(k, \mu_l) $ that can be solved on the user side and the AP side, respectively, more details are in \cite{9064520}  
\section{\uppercase{Performance evaluation}}
An indoor environment is considered with dimensions 5m $ \times  $ 5m $ \times $ 3m. On the ceiling, $ L =16 $ optical APs are deployed  to serve $ K $ users distributed on the receiving plane. Moreover, a WiFi AP is considered to provide uplink transmission as well as serving  users who penalize the sum rate of the optical wireless system. In this work, each user is equipped with a reconfigurable detector that contains multiple photodiodes in order to provide linearly independent channel responses from the whole set of the transmitters. In addition, each user has an RF antenna given the fact that some users might be served by the WiFi AP in accordance to the solution of the optimization problem in  \eqref{eq:ML}. The rest of the simulation parameters are listed in Table I.    

In Fig.\ref{rate_users}, the sum rate of the hybrid network is shown versus the number of users considering different values of the laser beam waist. Recall that the optimization problem formulated in this work  can be solved directly considering equation   \eqref{Lag}, i.e., a decentralized algorithm, due to its practicality where  a sub-optimal solution close to the solution of the centralized algorithm can be obtained with low complexity. The figure shows that the sum rate increases with the number of users where each user experiences a different channel gain. Note that, the hybrid network achieves higher sum rate compared with the standalone  optical wireless network, in which the WiFi AP provides only uplink transmission, and it is not considered in the optimization problem that maximizes  the sum rate of the users in  downlink transmission. Furthermore, the sum rates of all the scenarios increase with increase in the beam waist  of the laser due to the fact that the received power of each user increases considerably with the beam waist  $ W_{0} $ as more power is focused towards the users.  

Finally, Fig. \ref{cdf} shows the cumulative distribution function (CDF) of  the sum rate in the hybrid network implementing BIA or ZF schemes to align multi-user interference. It is shown that BIA is superior to ZF and more suitable for the optical wireless system due to the fact that BIA naturally satisfies the non-negativity of the optical signal, and therefore, it helps in  avoiding the need for applying a DC bias current to the optical signal, which might cause clipping distortion to useful information. It is worth mentioning that BIA suffers  data rate degradation as the number of users increases in the network due to channel coherence requirements and noise enhancement. However, in the case of  the hybrid network, some of the users negatively impacting the sum rate within the coverage of each optical AP can be moved to the WiFi system such that the overall sum rate of the hybrid network is maximized.   
\section{\uppercase{Conclusions}}
\label{sec:conclusion}
In this paper, the performance of an optical wireless system using IR lasers is  evaluated taking into consideration the existence of a WiFi system. First, the system model composed of  multiple IR lasers on the ceiling serving multiple users is defined. In addition, a WiFi system is considered to provide uplink transmission and serve users that overload the optical wireless system. Then, two transmission schemes are defined for the optical wireless system,  ZF and BIA, to derive the achievable user rates. After that,  an optimization problem is formulated  to jointly find the optimal user association and resource allocation that maximize  the overall sum rate of the hybrid network. The optimization problem is  a MINLP complex problem, which is difficult  to solve. Therefore, a decentralized  algorithm  is proposed  via  Lagrangian multipliers where the main problem is divided into sub-problems that can be solved separately. The results show significant enhancement in the sum rate of the hybrid optical/RF network after performing the optimization problem that considers all the APs in the environment including the WiFi system.    

\section{\uppercase{Future directions}}
IR laser-based optical wireless communication  can unlock  data rate speeds  in a range of Tbps. However, crucial   wireless communication issues ranging  from the physical layer, the data link layer and the network layer must be addressed including  transmitter and receiver designs, developing highly efficient interference management schemes and  optimal resource allocation approaches and  backhaul network designs, etc. In general, these issues might lead to formulating complex optimization problems that are difficult to solve. Note that, some optimization tools are available to solve  such complex optimization problems. However, practical algorithms  are highly required especially when real-time scenarios are concerned.
   
In recent years, machine learning (ML) techniques have been studied  for solving  NP-hard optimization problems with different contexts \cite{8743390,8970161}. For instance, some of the previous works in the literature related to the optimization problem  formulated in this paper  \cite{9566383,9159585,9136703} consider the implementation of  reinforcement learning (RL) to solve various optimization problems in optical and RF systems where RL can interact with   an environment to learn an optimal policy that makes right decisions. Specifically, in \cite{9566383}, reinforcement learning is  used for assigning each user to an exclusive wavelength in a WDMA-based optical wireless network. In \cite{9159585},  a deep Q-network (DQN) learning-based algorithm is proposed for solving an optimization problem that aims to  maximize the sum rate of a hybrid optical/RF network through allocating power and bandwidth in addition to user association. In \cite{9136703}, a RL-based load balancing approach is proposed for maximizing the sum rate of users in a hybrid LiFi/ WiFi network. It is shown that ML techniques can provide sub-optimal solutions while avoiding complexity. However, conventional ML techniques are not suitable in terms of providing immediate solutions  in large size  optical  wireless networks. As future work, deep learning can be used for solving various optimization problems as it is expected to be superior to ML techniques where a learning process is preformed over a data set generated  in an offline phase to provide sub-optimal solutions in a real time phase.   
\section*{\uppercase{Acknowledgements}}
This work  has been supported in by the Engineering and Physical Sciences Research Council (EPSRC), in part by the INTERNET project under Grant EP/H040536/1, and in part by the STAR project under Grant EP/K016873/1 and in part by the TOWS project under Grant EP/S016570/1. All data are provided in full in the results section of this paper.

\bibliographystyle{apalike}
{\small
\bibliography{text}}

\end{document}